\documentclass[aps,prb,twocolumn,superscriptaddress,groupedaddress,a4paper]{revtex4}  
\pdfoutput=1
\usepackage{graphicx}
\usepackage{dcolumn}
\usepackage{bm}
\usepackage{amssymb}
\usepackage{amsmath}
\usepackage{subfigure}
\usepackage{color}
\begin{document}

\widetext
\title{12-band $\textbf{k}\cdot\textbf{p}$ model for dilute bismide alloys of (In)GaAs derived from supercell calculations}


\author{Christopher A. Broderick}
\email{chris.broderick@tyndall.ie} 
\affiliation{Tyndall National Institute, Lee Maltings, Dyke Parade, Cork, Ireland}
\affiliation{Department of Physics, University College Cork, Cork, Ireland}

\author{Muhammad Usman}
\affiliation{Tyndall National Institute, Lee Maltings, Dyke Parade, Cork, Ireland}

\author{Eoin P. O'Reilly}
\affiliation{Tyndall National Institute, Lee Maltings, Dyke Parade, Cork, Ireland}
\affiliation{Department of Physics, University College Cork, Cork, Ireland}

\vskip 0.25cm

\date{\today}


\begin{abstract}

Incorporation of bismuth (Bi) in dilute quantities in (In)GaAs has been shown to lead to unique electronic properties that can in principle be exploited for the design of high efficiency telecomm lasers. This motivates the development of simple models of the electronic structure of these dilute bismide alloys, which can be used to evaluate their potential as a candidate material system for optical applications. Here, we begin by using detailed calculations based on an $sp^{3}s^{*}$ tight-binding model of (In)GaBi$_{x}$As$_{1-x}$ to verify the presence of a valence band-anticrossing interaction in these alloys. Based on the tight-binding model the derivation of a 12-band $\textbf{k}\cdot\textbf{p}$ Hamiltonian for dilute bismide alloys is outlined. We show that the band structure obtained from the 12-band model is in excellent agreement with full tight-binding supercell calculations. Finally, we apply the 12-band model to In$_{0.53}$Ga$_{0.47}$Bi$_{x}$As$_{1-x}$ and compare the calculated variation of the band gap and spin-orbit-splitting to a variety of spectroscopic measurements performed on a series of MBE-grown In$_{0.53}$Ga$_{0.47}$Bi$_{x}$As$_{1-x}$/InP layers.

\end{abstract}


\maketitle


\section{Introduction}
\label{sec:introduction}

Incorporation of dilute concentrations of bismuth (Bi) in III-V alloys such as GaAs and InGaAs to form the dilute bismide alloys GaBi$_{x}$As$_{1-x}$ and In$_{1-y}$Ga$_{y}$Bi$_{x}$As$_{1-x}$ has been experimentally found to rapidly reduce the band gap energy ($E_{g}$) by up to 90 meV \cite{Tixier_APL_2005,Alberi_PRB_2007} and 60 meV \cite{Petropoulos_APL_2011,Marko_APL_2012}, respectively, per \% Bi for $x \lesssim$ 4 -- 5\%. Additionally, experiments have also revealed the presence of a large bowing of the spin-orbit-splitting energy ($\Delta_{\scalebox{0.7}{\textrm{SO}}}$) in these alloys, which increases strongly with increasing Bi composition \cite{Fluegel_PRL_2006}. It has recently been demonstrated experimentally \cite{Batool_JAP_2012} and verified theoretically \cite{Usman_PRB_2011} that these two properties lead, at sufficiently large $x$ ($\sim$ 10\%), to the onset of an $E_{g} < \Delta_{\scalebox{0.7}{\textrm{SO}}}$ regime in GaBi$_{x}$As$_{1-x}$.

These characteristics of the dilute bismide alloys are highly attractive for potential device applications as they open up the possibility to suppress the CHSH Auger recombination mechanism, a non-radiative recombination process that severely degrades the efficiency and temperature stability of conventional InP-based photonic devices operating in the telecomm wavelength range (1.3 -- 1.5 $\mu$m) \cite{Sweeney_PSSB_1999,Higashi_JSTQE_1999,Sweeney_ICTON_2011,Broderick_SST_2012}. This motivates the development of simple models of the electronic structure of (In)GaBiAs alloys, which can be applied to study the characteristics of dilute bismide-based devices.

Since Bi is significantly larger and more electropositive than As, it should be expected that Bi behaves as an impurity in (In)GaAs and that, as a result of this, it should introduce Bi-related defect states close in energy to the (In)GaAs valence band edge (VBE). Bi has been experimentally shown to form bound states in GaP:Bi, lying roughly 0.1 eV above the GaP VBE \cite{Trumbore_APL_1966}. Similarly, the formation of Bi-related resonant states lying below the GaAs VBE has been demonstrated by theoretical calculations of the valence band structure of GaBi$_{x}$As$_{1-x}$ \cite{Usman_PRB_2011,Zhang_PRB_2005,Broderick_12band_2012}. 

The large bowing of $E_{g}$ and $\Delta_{\scalebox{0.7}{\textrm{SO}}}$ in (In)GaBi$_{x}$As$_{1-x}$ and related alloys has been explained previously in terms of a band-anticrossing (BAC) interaction  between the GaAs VBE and Bi-related impurity states lying below it in energy \cite{Alberi_PRB_2007}, similar to that occuring in the conduction band in dilute nitride alloys. This simple phenomenological model has been successful in reproducing the measured variation of $E_{g}$ and $\Delta_{\scalebox{0.7}{\textrm{SO}}}$ with Bi composition in GaBi$_{x}$As$_{1-x}$. The initial models presented relied upon fitting the energy of the Bi-related states ($E_{\scalebox{0.7}{\textrm{Bi}}}$) and the BAC coupling parameter ($\beta$) to the experimentally measured variation of $E_{g}$ and $\Delta_{\scalebox{0.7}{\textrm{SO}}}$ with Bi composition, while also treating the GaBi$_{x}$As$_{1-x}$ conduction band offset with respect to GaAs as an additional free parameter. For example, such an approach has been applied to GaBi$_{x}$As$_{1-x}$ and In$_{1-y}$Ga$_{y}$Bi$_{x}$As$_{1-x}$ alloys in Refs.~\onlinecite{Alberi_PRB_2007} and~\onlinecite{Petropoulos_APL_2011}.

Based on an $sp^{3}s^{*}$ tight-binding model, calculations on large, disordered GaBi$_{x}$As$_{1-x}$ alloy supercells have verified the presence of BAC-like behaviour in the valence band, while also reproducing the measured variation of $E_{g}$ and $\Delta_{\scalebox{0.7}{\textrm{SO}}}$ with $x$ over a large composition range (up to $x \sim$ 13$\%$) \cite{Usman_PRB_2011}. Since the tight-binding model employs a basis of localised atomic orbitals, it is well suited to probe the character of states associated with an isolated impurity in large supercells \cite{Reilly_JPCS_2010}. Detailed calculations on ordered GaBi$_{x}$P$_{1-x}$ and GaBi$_{x}$As$_{1-x}$ supercells using the tight-binding model have confirmed the presence of the bound and resonant Bi-related states in GaP:Bi and GaAs:Bi, respectively \cite{Usman_PRB_2011,Broderick_12band_2012}. In this approach, by examining the character of the Bi-related states, we can directly derive (i) the BAC parameters $E_{\scalebox{0.7}{\textrm{Bi}}}$ and $\beta$, and (ii) the composition dependence of the band offsets, in a given Bi-containing supercell \cite{Usman_PRB_2011,Broderick_12band_2012}. This then eliminates the need to fit these quantities to the results of experimental measurements.

In this article we apply our tight-binding model to ordered GaBi$_{x}$As$_{1-x}$ and InBi$_{x}$As$_{1-x}$ supercells and outline the derivation of a 12-band $\textbf{k}\cdot\textbf{p}$ Hamiltonian for (In)GaBi$_{x}$As$_{1-x}$ (the full details of which can be found in Ref.~\onlinecite{Broderick_12band_2012}). The 12-band model, which includes BAC interactions between the extended states of the host (In)GaAs matrix and Bi-related resonant states lying below the host matrix VBE, is shown to be in excellent agreement with full tight-binding calculations of the band structure in Bi-containing supercells, verifying the validity of the BAC approach to dilute bismide band structure. We demonstrate this by applying the 12-band model to In$_{1-y}$Ga$_{y}$Bi$_{x}$As$_{1-x}$ alloys and comparing the results of our calculations to the available experimental data for this new material system.

The remainder of the article is organised as follows: In Section~\ref{sec:tight_binding_model} we give a brief overview of the tight-binding model used to study dilute bismide alloys. In Section~\ref{sec:ordered_supercells} this model is applied to ordered GaBi$_{x}$As$_{1-x}$ and InBi$_{x}$As$_{1-x}$ supercells and the parameters of the BAC model are calculated explicitly. Then, in Section~\ref{sec:kdotp_models}, we use the insight gained from the tight-binding calculations to derive an appropriate 12-band $\textbf{k}\cdot\textbf{p}$ Hamiltonian for (In)GaBi$_{x}$As$_{1-x}$. Finally, in Section~\ref{sec:comparison_with_experiment} we verify the validity of the BAC approach by applying the $\textbf{k}\cdot\textbf{p}$ models of Section~\ref{sec:kdotp_models} to the study of In$_{0.53}$Ga$_{0.47}$Bi$_{x}$As$_{1-x}$ alloys.


\section{Tight-binding model}
\label{sec:tight_binding_model}

We presented in Ref.~\onlinecite{Usman_PRB_2011} a nearest-neighbour $sp^{3}s^{*}$ tight-binding Hamiltonian, including spin-orbit-coupling, for GaBi$_{x}$As$_{1-x}$. Here we combine this Hamiltonian with $sp^{3}s^{*}$ Hamiltonians for InAs and InBi to give an $sp^{3}s^{*}$ Hamiltonian for the quaternary dilute bismide alloy In$_{1-y}$Ga$_{y}$Bi$_{x}$As$_{1-x}$.

In our tight-binding model we begin by relaxing the atomic positions in a given (In)GaBi$_{x}$As$_{1-x}$ supercell to their lowest energy configuration using a valence force field model based on the Keating potential \cite{Keating_PR_1966,Lazarenkov_APL_2004}.

The on-site energies of the relaxed supercell Hamiltonian are taken to depend on the overall neighbour environment and the inter-atomic interaction energies are taken to vary with the relaxed nearest-neighbour bond length $d$ as $\left( \frac{d_{0}}{d} \right)^{\eta}$, where $d_{0}$ is the equilibrium bond length in the equivalent binary compound, and $\eta$ is a dimensionless scaling parameter (the magnitude of which depends on the type of interaction being considered). Full details of the tight-binding model can be found in Ref.~\onlinecite{Usman_PRB_2011}.


\section{B\lowercase{i} impurity states in ordered (I\lowercase{n})G\lowercase{a}B\lowercase{i}$_{x}$A\lowercase{s}$_{1-x}$ supercells}
\label{sec:ordered_supercells}

We examine the electronic structure of ordered GaBi$_{x}$As$_{1-x}$ and InBi$_{x}$As$_{1-x}$ crystals by inserting a single substitutional Bi atom in a series of cubic Y$_{M}$Bi$_{1}$As$_{M-1}$ (Y = Ga, In) supercells containing a total of $2M = 8N^{3}$ atoms, for $2 \leq N \leq 8$.

In Ref.~\onlinecite{Usman_PRB_2011} we showed that the spectrum of fractional $\Gamma$ character for a given alloy supercell, $G_{\Gamma} (E)$, is a useful tool for examining the evolution of the electronic structure of impurity containing alloys, where $G_{\Gamma}  (E)$ is defined as the projection of the host matrix (GaAs or InAs) band edge states ($\vert \psi_{n,0} \rangle$) onto the full spectrum of levels in the alloy (YBi$_{x}$As$_{1-x}$) supercells, $\left\lbrace E_{i}, \vert \psi_{i,1} \rangle \right\rbrace$:

\begin{equation}
	\label{eq:Gamma_character}
	G_{\Gamma} \left( E \right) = \sum_{i} \sum_{n=1}^{g(E_{n})}   \vert \langle \psi_{i,1} \vert \psi_{n,0} \rangle \vert ^{2} \; \delta \left( E_{i} - E \right)
\end{equation}

\noindent
where  $g(E_{n})$ is the degeneracy of the host band having energy $E_{n}$ at the $\Gamma$-point in the Brillouin zone and $\delta \left( E_{i} - E \right)$ is a Dirac delta function centred at energy $E_{i}$.

Calculations of $G_{\Gamma} (E)$ in Y$_{M}$Bi$_{1}$As$_{M-1}$ supercells show mixing between the host YAs VBE states and Bi-related states lying below it in energy, consistent with the presence of a BAC interaction in the YBi$_{x}$As$_{1-x}$ valence band \cite{Broderick_12band_2012}.

In order to confirm this, we examine the BAC model explicitly for each Bi-containing supercell by constructing the Bi-related impurity state $\vert \psi_{\scalebox{0.7}{\textrm{Bi}}} \rangle$ associated with the substitutional Bi atom present in each supercell. In the BAC model, the alloy VBE states $\vert \psi_{1,i} \rangle$ are a linear combination of the four host VBE states $\vert \psi_{0,n} \rangle$ (at energy $E_{v,0}$) and the Bi-related defect states $\vert \psi_{\scalebox{0.7}{\textrm{Bi}},i} \rangle$:

\begin{equation}
	\label{eq:Bi_defect_state}
	\vert \psi_{\scalebox{0.7}{\textrm{Bi}},i} \rangle = \frac{\vert \psi_{1,i} \rangle - \sum_{n} \vert \psi_{0,n} \rangle \langle \psi_{0,n} \vert \psi_{1,i} \rangle}{\sqrt{1 - \sum_{n} \vert \langle \psi_{0,n} \vert \psi_{1,i} \rangle \vert^{2}}}
\end{equation}

\noindent
where the indices $n$ and $i$ run over the four-fold degenerate states of the Y$_{M}$As$_{M}$ and Y$_{M}$Bi$_{1}$As$_{M-1}$ VBEs, respectively \cite{Usman_PRB_2011}.

From Eq.~\eqref{eq:Bi_defect_state} we can calculate the energy of the Bi states and the strength of their interaction with the host VBE using the full $sp^{3}s^{*}$ Hamiltonian of the Y$_{M}$Bi$_{1}$As$_{M-1}$ supercell, $\widehat{H}$, to find:

\begin{equation}
	\label{eq:Bi_state_energy}
	E_{\scalebox{0.7}{\textrm{Bi}},i} = \langle \psi_{\scalebox{0.7}{\textrm{Bi}},i} \vert \widehat{H} \vert \psi_{\scalebox{0.7}{\textrm{Bi}},i} \rangle
\end{equation}

\begin{equation}
	\label{eq:Bi_state_interaction}
	V_{\scalebox{0.7}{\textrm{Bi}},i} = \langle \psi_{\scalebox{0.7}{\textrm{Bi}},i} \vert \widehat{H} \vert \psi_{v0,i} \rangle = \beta \sqrt{x}
\end{equation}

\noindent
where $x = M^{-1}$ is the Bi composition, $\beta$ is the BAC interaction parameter, and $\vert \psi_{v0,i} \rangle$ is the host VBE state with which $\vert \psi_{\scalebox{0.7}{\textrm{Bi}},i} \rangle$ interacts \cite{Usman_PRB_2011}.

Our results show that a substitutional Bi atom in Ga$_{M}$Bi$_{1}$As$_{M-1}$ or in In$_{M}$Bi$_{1}$As$_{M-1}$ forms, a set of four-fold degenerate resonant states at energy $E_{\scalebox{0.7}{\textrm{Bi}}}$, which are highly localised about the Bi site, and that the strength of the interaction between these states and the host VBE varies with Bi composition $x$ as $V_{\scalebox{0.7}{\textrm{Bi}}} = \beta \sqrt{x}$ for large supercells.

Figure 1 shows the calculated values of the energy separation between the host VBE and Bi resonant state, $\Delta E_{\scalebox{0.7}{\textrm{Bi}}} = E_{\scalebox{0.7}{\textrm{Bi}}} - E_{v,0}$, as well as the BAC coupling parameter $\beta$, as a function of supercell size for a series of ordered Y$_{M}$Bi$_{1}$As$_{M-1}$ (Y = Ga, In) supercells. We observe a strong variation of $\Delta E_{\scalebox{0.7}{\textrm{Bi}}}$ and $\beta$ with Bi composition at larger $x$, with the trends stabilising as $x$ approaches the dilute doping limit in the larger ($\gtrsim$ 2000 atom) supercells \cite{Usman_PRB_2011}. This is primarily due to the folding of the host valence band states back to the $\Gamma$ point as the supercell size is increased, which makes more states available to construct $\vert \psi_{\scalebox{0.7}{\textrm{Bi}},i} \rangle$ in the larger supercells. Therefore, we conclude that one should refer to large supercell calculations when examining the behaviour of a substitutional Bi impurity in (In)GaAs alloys. A detailed discussion of this aspect of the calculations and its consequences for the interpretation of the BAC model as applied to the valence band structure of dilute bismide alloys can be found in Ref.~\onlinecite{Broderick_12band_2012}.


\begin{table}[t!]
		\caption{\label{tab:BAC_parameters} Calculated energy of the Bi impurity state relative to the host matrix VBE ($\Delta E_{\protect\scalebox{0.7}{\textrm{Bi}}}$) and BAC coupling parameter ($\beta$) for substitutional Bi incorporation in 64-atom ($N = 2$) and 4096-atom ($N = 8$) Ga$_{M}$Bi$_{1}$As$_{M-1}$ and In$_{M}$Bi$_{1}$As$_{M-1}$ supercells. The 64-atom values are given in parentheses.}
	\begin{ruledtabular}
		\begin{center}
			\begin{tabular}{ccc}
				                                          & GaAs           & InAs          \\
				\hline
				$\Delta E_{\scalebox{0.7}{\rm{Bi}}}$ (eV) & (-0.60) -0.18  & (-0.67) -0.22 \\
				$\beta$ (eV)                              & (1.42)   1.13  & (1.19)   0.92 \\
			\end{tabular}
		\end{center}
	\end{ruledtabular}
\end{table}


\begin{figure}[t!]
	\centering
	\includegraphics*[width=\linewidth,height=1.1\linewidth]{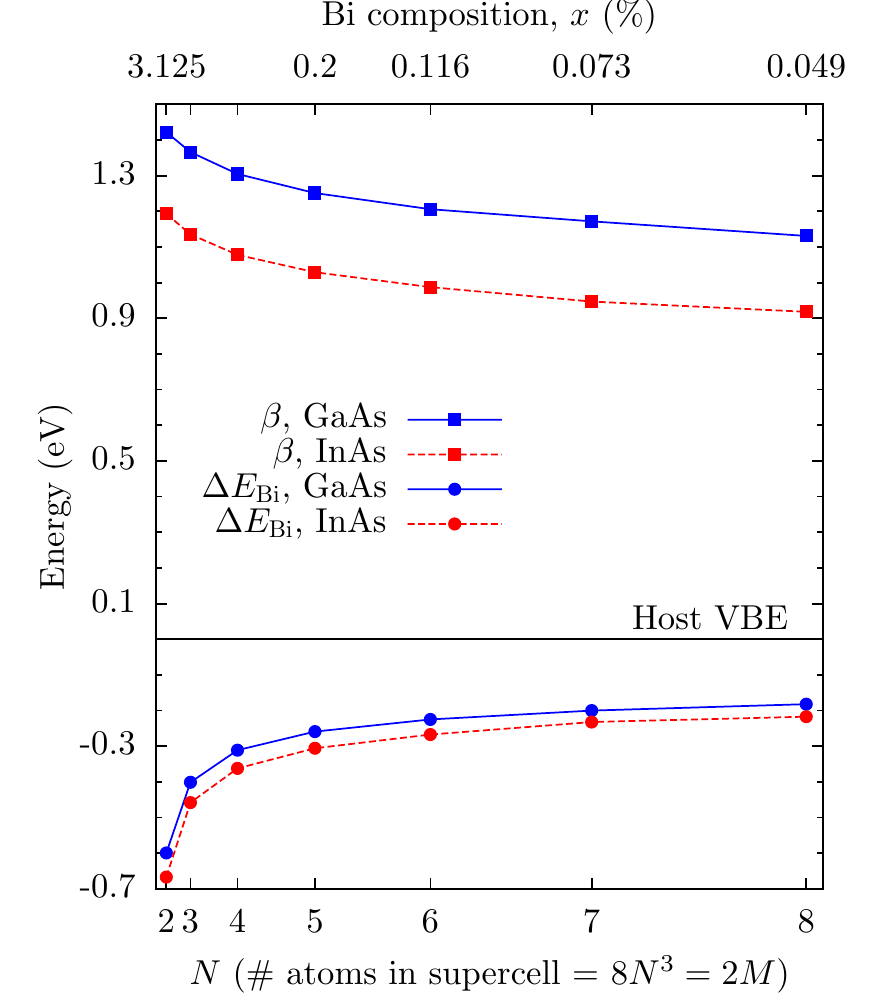}
	\caption{Calculated separation between the Bi impurity state and host VBE, $\Delta E_{\protect\scalebox{0.7}{\textrm{Bi}}}$ (circles), and BAC coupling parameter, $\beta$ (squares), for substitutional Bi incorporation in a series of ordered, cubic Y$_{M}$Bi$_{1}$As$_{M-1}$ (Y = Ga, In) supercells. As a guide to the eye, solid (dashed) lines have been inserted to highlight the evolution of the BAC parameters with Bi composition in the Ga$_{M}$Bi$_{1}$As$_{M-1}$ (In$_{M}$Bi$_{1}$As$_{M-1}$) cases. The zero of energy is taken in each case at the host VBE and is denoted by a horizontal line at 0 eV.}
	\label{impurity_states}
\end{figure}

It is interesting to note here the similar character of the Bi-related states in Ga$_{M}$Bi$_{1}$As$_{M-1}$ and in In$_{M}$Bi$_{1}$As$_{M-1}$ -- the impurity states associated with isolated Bi impurities in GaAs and InAs lie within 0.1 eV of one another in all supercells considered. This is in stark contrast to the case of a substitutional nitrogen (N) atom in GaAs or in InAs, where the N-related impurity states lie approximately 0.19 eV and 0.97 eV above the host conduction band edge respectively \cite{Reilly_SST_2009}. In the dilute nitride case this different behavior arises primarily due to differences in the GaAs and InAs conduction band structures \cite{Lindsay_PB_2003}. Since Bi primarily affects the valence band in (In)GaBi$_{x}$As$_{1-x}$, we then attribute the similarity of the Bi-related states in GaBi$_{x}$As$_{1-x}$ and InBi$_{x}$As$_{1-x}$ to the relative similarity of the valence band structure of GaAs and InAs, including a very similar spin-orbit-splitting energy of 340 meV and 380 meV respectively. By contrast the spin-orbit-splitting energy is only 80 meV in GaP. This difference in band structure, as well as the greater difference in size and electronegativity between P and Bi compared to that between As and Bi then leads to Bi introducing a defect state about 0.1 eV above the valence band maximum in GaP~\cite{Usman_PRB_2011,Trumbore_APL_1966}.

We list in Table~\ref{tab:BAC_parameters} the calculated values of $\Delta E_{\scalebox{0.7}{\textrm{Bi}}}$ and $\beta$ in the Y$_{32}$Bi$_{1}$As$_{31}$ ($N = $ 2) and Y$_{2048}$Bi$_{1}$As$_{2047}$ ($N = $ 8) supercells. Our calculations predict that a substitutional Bi impurity produces a resonant state lying approximately 0.18 eV (0.22 eV) below the GaAs (InAs) VBE in the dilute doping limit.

Using these detailed supercell calculations to further inform our analysis, we derive in the next section a 12-band $\textbf{k}\cdot\textbf{p}$ Hamiltonian to describe the band structure of ordered GaBi$_{x}$As$_{1-x}$ and InBi$_{x}$As$_{1-x}$ crystals.


\section{12-band $\textbf{k}\cdot\textbf{p}$ models of (I\lowercase{n})G\lowercase{a}B\lowercase{i}$_{x}$A\lowercase{s}$_{1-x}$}
\label{sec:kdotp_models}


\begin{figure*}[t!]
	\hspace{-0.20cm} \subfigure{ \includegraphics[width=0.42\textwidth]{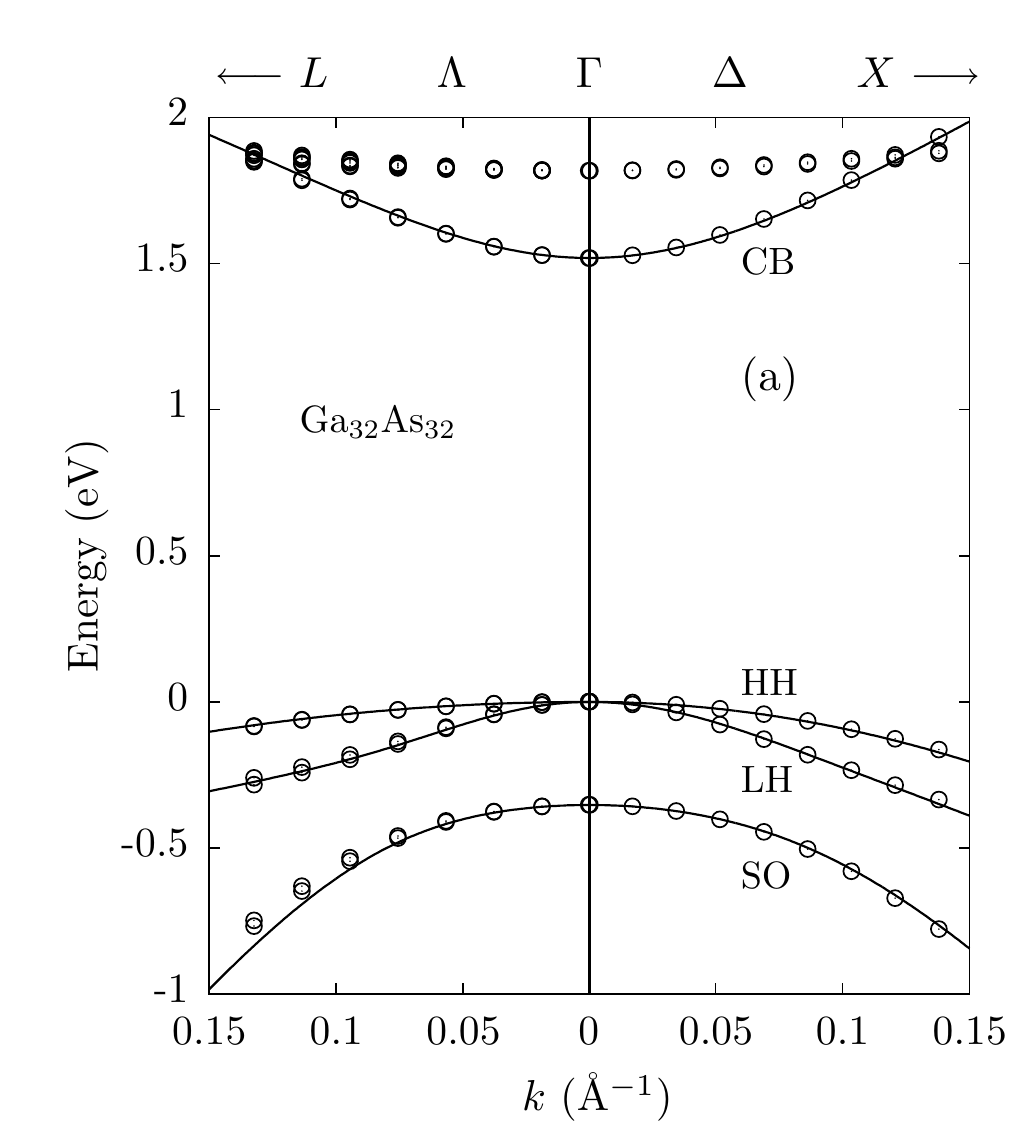} \label{fig:Ga32As32}    }
	\hspace{-0.20cm} \subfigure{ \includegraphics[width=0.42\textwidth]{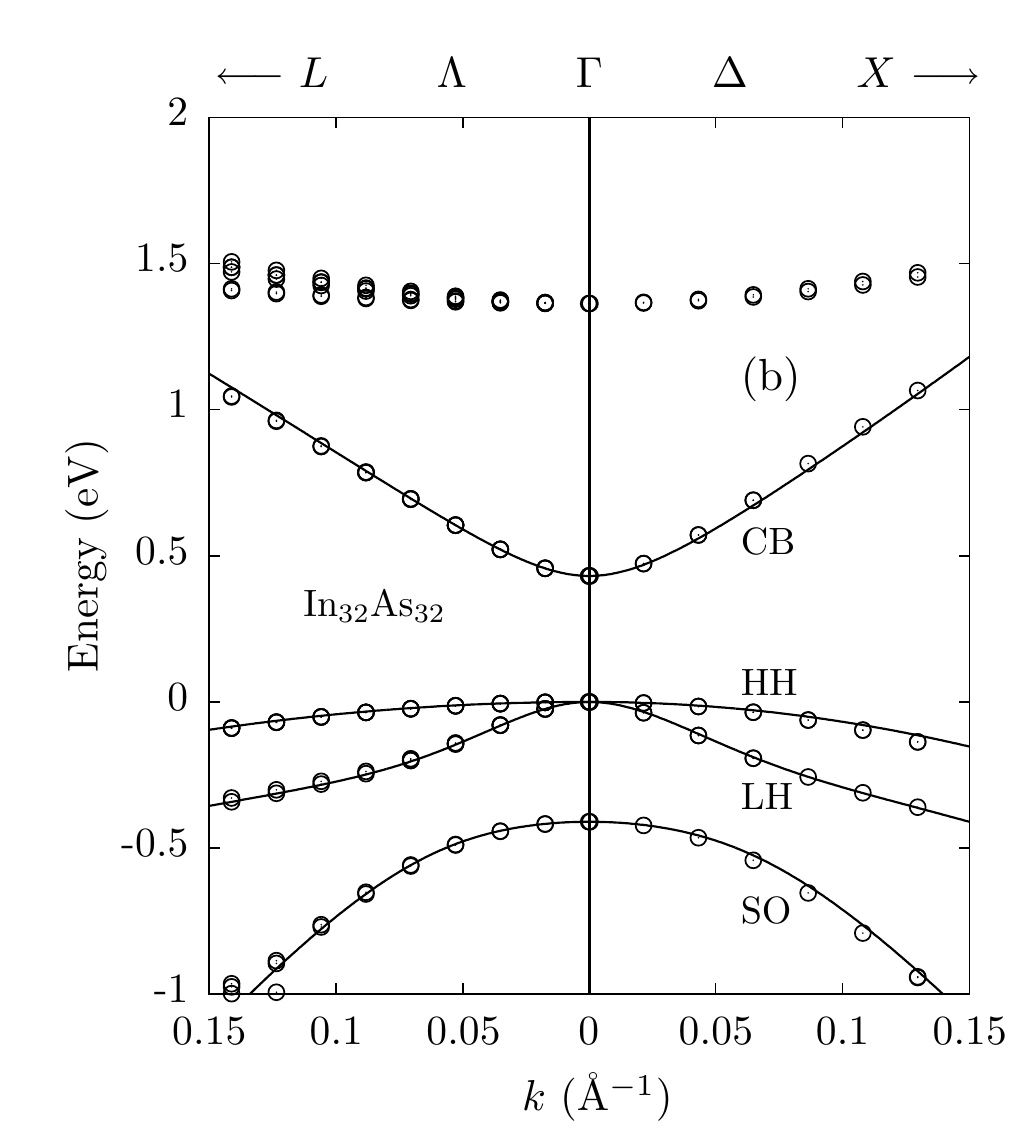} \label{fig:In32As32}    }
	\hspace{-0.80cm} \subfigure{ \includegraphics[width=0.42\textwidth]{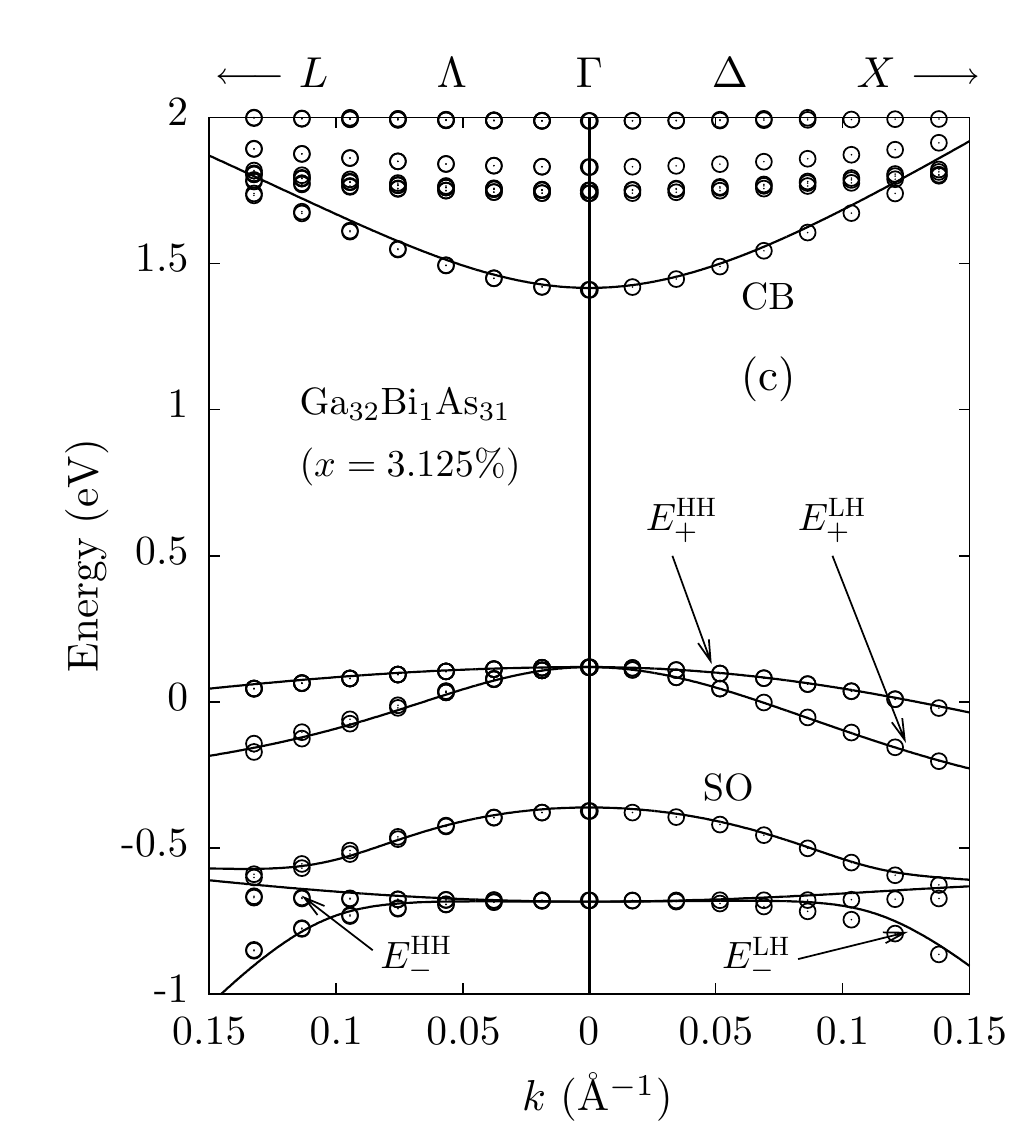} \label{fig:Ga32Bi1As31} }
	\hspace{-0.20cm} \subfigure{ \includegraphics[width=0.42\textwidth]{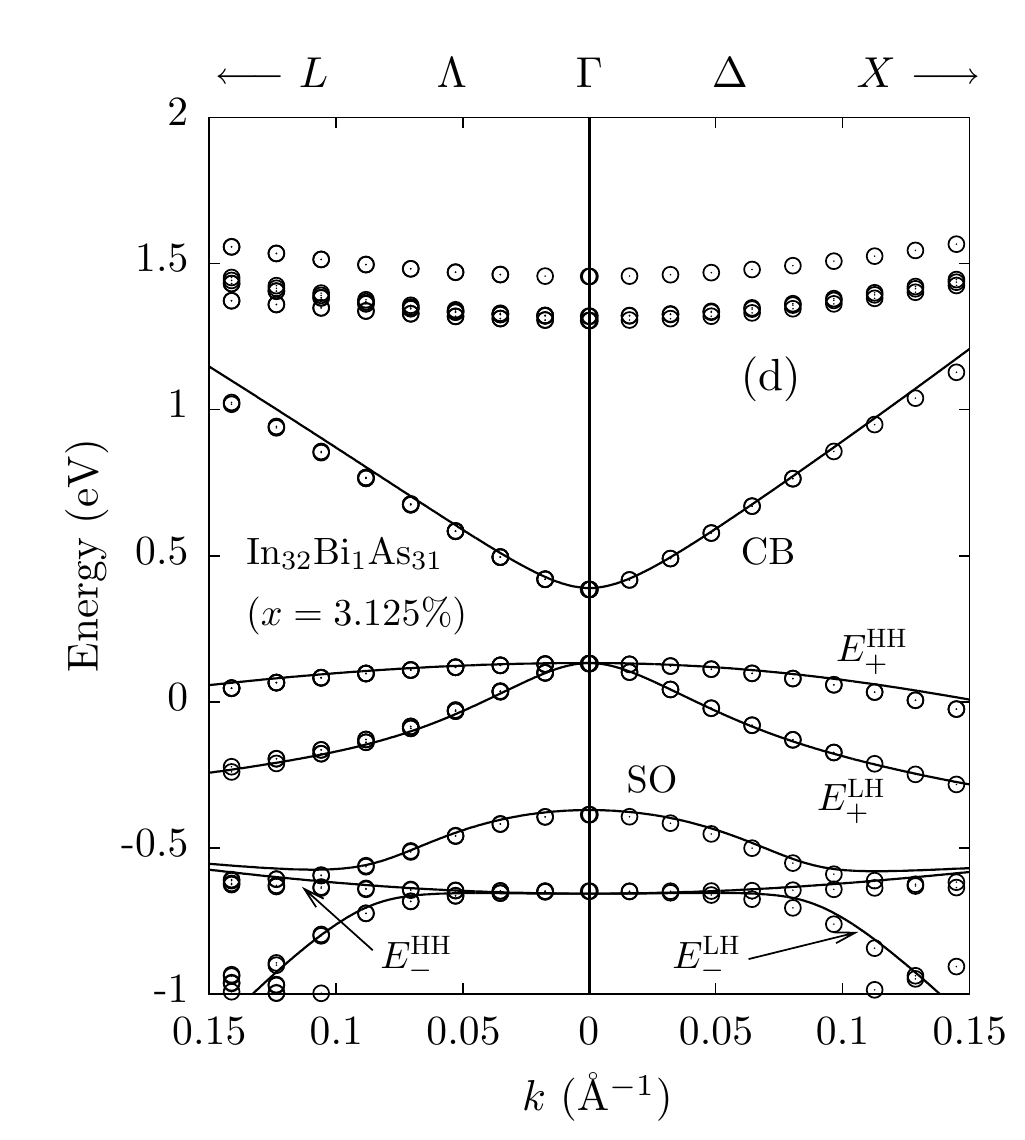} \label{fig:In32Bi1As31} }
	\caption{Band structure of (a) Ga$_{32}$As$_{32}$, and (b) In$_{32}$As$_{32}$, calculated using $sp^{3}s^{*}$ tight-binding (open circles), and 8-band $\textbf{k}\cdot\textbf{p}$ (solid lines) Hamiltonians. Band structure of (c) Ga$_{32}$Bi$_{1}$As$_{31}$, and (d) In$_{32}$Bi$_{1}$As$_{31}$ ($x =$ 3.125\%) calculated using $sp^{3}s^{*}$ tight-binding (open circles), and 12-band $\textbf{k}\cdot\textbf{p}$ (solid lines) Hamiltonians. The band structures are shown here along the $\Lambda$ and $\Delta$ directions, close to the $\Gamma$ point in the Brillouin zone.}
\end{figure*}

As we have seen, detailed tight-binding calculations on ordered supercells show that (i) Bi incorporation in (In)GaAs should be treated as impurity-like, with substitutional incorporation resulting in a set of four-fold Bi-related impurity states lying below the host VBE, and (ii) that these states interact with the extended host VBE states through a Bi composition-dependent BAC interaction.

To arrive at an appropriate $\textbf{k}\cdot\textbf{p}$ Hamiltonian for YBi$_{x}$As$_{1-x}$ (Y = Ga, In) we begin by noting that large supercell calculations on disordered GaBi$_{x}$As$_{1-x}$ and InBi$_{x}$As$_{1-x}$ alloys indicate a conventional alloy evolution of the conduction and spin-split-off bands with $x$, with for example, an approximately linear decrease in the energy of the GaBi$_{x}$As$_{1-x}$ conduction band minimum at $\Gamma$ of about 28 meV per \% Bi \cite{Usman_PRB_2011}.

Beginning with an 8-band $\textbf{k}\cdot\textbf{p}$ Hamiltonian for YAs (Y = Ga, In) and combining this with the calculations presented in Section~\ref{sec:ordered_supercells}, we can then derive a 12-band $\textbf{k}\cdot\textbf{p}$ model for (In)GaBi$_{x}$As$_{1-x}$. The 12-band model is obtained by augmenting the 8-band basis of host matrix band edge Bloch states through the inclusion of a set of four-fold degenerate Bi-related basis states at energy $E_{\scalebox{0.7}{\textrm{Bi}}}$, below the VBE. The Bi-related states can be selected to have the same symmetry as the heavy- and light-hole (HH and LH) host states, and the resulting pairs of two-fold degenerate Bi states then interact with the host HH and LH states via the Bi composition dependent BAC matrix elements $V_{\scalebox{0.7}{\textrm{Bi}}} = \beta \sqrt{x}$. Full details of the 12-band model are given in Ref.~\onlinecite{Broderick_12band_2012}. 

We note that the dilute bismide $\textbf{k}\cdot\textbf{p}$ Hamiltonian presented here does not include a BAC interaction with an additional (two-fold degenerate) Bi-related spin-split-off impurity level lying deeper in the valence band, as was considered in Ref.~\onlinecite{Alberi_PRB_2007}. Our previous calculations in Ref.~\onlinecite{Usman_PRB_2011} indicate that in GaBi$_{x}$As$_{1-x}$ the energy of the spin-split-off band edge decreases with increasing Bi composition $x$. A detailed analysis showed that the evolution of the spin-split-off band with increasing Bi composition in GaBi$_{x}$As$_{1-x}$ can in fact be described using a conventional alloy model, and no evidence was found for additional Bi-related states lying lower in the valence band. Tight-binding calculations show the evolution of the spin-split-off band in InBi$_{x}$As$_{1-x}$ to be similar to that in GaBi$_{x}$As$_{1-x}$, so that it is then appropriate to treat the impact of Bi using a 12-band Hamiltonian, without the need to include the two addiitonal Bi-related states presented in the 14-band model of Ref.~\onlinecite{Alberi_PRB_2007}.

In order to compare the results of the 12-band model with the full tight-binding (FTB) calculations we begin by first establishing 8-band $\textbf{k}\cdot\textbf{p}$ models for GaAs and InAs. The zone centre (In)GaAs states obtained from the FTB calculation at $\Gamma$ provide the basis states to parametrise the 8-band GaAs and InAs Hamiltonians directly from their respective $sp^{3}s^{*}$ Hamiltonians \cite{Lindsay_SSC_2003}. The 8-band $\textbf{k}\cdot\textbf{p}$ models obtained from this approach for GaAs and InAs are compared to the FTB band structures in Figures~\ref{fig:Ga32As32} and~\ref{fig:In32As32}, where we observe excellent agreement between both sets of calculations in the vicinity of the $\Gamma$ point.   

Using these 8-band models as starting points, we construct the 12-band Hamiltonian for a given Bi-containing supercell by firstly using Eqs.~\eqref{eq:Bi_state_energy} and~\eqref{eq:Bi_state_interaction} to calculate $E_{\scalebox{0.7}{\textrm{Bi}}}$ and $\beta$ for the supercell. These parameters are then combined with the 8-band parameters of the corresponding Bi-free supercell as well as the Bi composition dependence of the band edge energies \cite{Broderick_12band_2012} to give the appropriate $\textbf{k}\cdot\textbf{p}$ Hamiltonian for the Y$_{M}$Bi$_{1}$As$_{M-1}$ supercell.

Figures~\ref{fig:Ga32Bi1As31} and~\ref{fig:In32Bi1As31} compare the band structures calculated using this approach with the FTB calculations of the band structure for the 64-atom, Ga$_{32}$Bi$_{1}$As$_{31}$ and In$_{32}$Bi$_{1}$As$_{31}$ supercells ($x =$ 3.125\%). The calculations demonstrate excellent agreement between the $sp^{3}s^{*}$ and 12-band models. The presence of a BAC interaction in the valence band is confirmed by the existence of tight-binding bands that correspond in both cases to the lower energy Bi-related impurity bands ($E_{-}$) of the 12-band Hamiltonian. These exemplar calculations confirm the validity of the BAC approach as applied to dilute bismide valence band structure.

It is, however, important to note that the direct correspondence observed between the Bi impurity bands in the calculations presented in Figures~\ref{fig:Ga32Bi1As31} and~\ref{fig:In32Bi1As31} occurs because the highest folded valence bands in the 64-atom Y$_{32}$As$_{32}$ (Y = Ga, In) supercells lie over 1 eV below the host VBE. This results in a low density of host valence states in the proximity of the Bi-related impurity states, when compared to larger supercells in which there is a higher density of folded host matrix states, leading to the Bi impurity forming a resonant state, which is spread over several of the host levels. This energy broadening of the lower-lying impurity bands accounts for the inability to resolve any Bi-related features lying below the VBE in spectroscopic measurements \cite{Usman_PRB_2011}.

In summary, the overall agreement of the 12-band $\textbf{k}\cdot\textbf{p}$ model with the FTB calculations confirms that the BAC model accurately describes the main effects of Bi on the band structure of ordered (In)GaBi$_{x}$As$_{1-x}$ structures. In the next section, we apply the 12-band $\textbf{k}\cdot\textbf{p}$ model to compare the calculated variations of the energy gap, E$_{g}$ and spin-orbit-splitting energy $\Delta_{\scalebox{0.7}{\textrm{SO}}}$ with spectroscopic measurements on a series of In$_{0.53}$Ga$_{0.47}$Bi$_{x}$As$_{1-x}$/InP epilayers.


\section{Application of the 12-band $\textbf{k}\cdot\textbf{p}$ model to I\lowercase{n}$_{1-y}$G\lowercase{a}$_{y}$B\lowercase{i}$_{x}$A\lowercase{s}$_{1-x}$ and comparison with experiment}
\label{sec:comparison_with_experiment}

We now extend the 12-band model outlined in Section~\ref{sec:kdotp_models} and Ref.~\onlinecite{Broderick_12band_2012} for GaBi$_{x}$As$_{1-x}$ and InBi$_{x}$As$_{1-x}$ to the quaternary dilute bismide alloy In$_{1-y}$Ga$_{y}$Bi$_{x}$As$_{1-x}$. Using the derived $\textbf{k}\cdot\textbf{p}$ Hamiltonian we demonstrate the validity of the approach outlined in Sections~\ref{sec:ordered_supercells} and~\ref{sec:kdotp_models} by comparing the calculated variations of the room temperature band gap and spin-orbit-splitting energies with Bi composition against spectroscopic measurements on a series of molecular beam epitaxy (MBE) grown In$_{0.53}$Ga$_{0.47}$Bi$_{x}$As$_{1-x}$ samples, grown pseudomorphically on InP:Fe substrates. Details of the sample growth and spectroscopic measurements can be found in Refs.~\onlinecite{Petropoulos_APL_2011} and~\onlinecite{Marko_APL_2012}.

In order to extend the 12-band models of Section~\ref{sec:kdotp_models} to the quaternary In$_{1-y}$Ga$_{y}$Bi$_{x}$As$_{1-x}$ alloy we linearly interpolate between the BAC parameters and Bi compositional dependence of the band offsets for GaBi$_{x}$As$_{1-x}$ and InBi$_{x}$As$_{1-x}$ derived from large supercell tight-binding calculations (cf. Table~\ref{tab:BAC_parameters} and Refs.~\onlinecite{Usman_PRB_2011} and \onlinecite{Broderick_12band_2012}). The In$_{1-y}$Ga$_{y}$Bi$_{x}$As$_{1-x}$ layer is taken as being uniformly strained onto an InP substrate. We use V\'{e}gard's law to interpolate between the lattice and elastic constants of the constituent binary compounds in order to obtain those of the quaternary alloy. We also use linear interpolation for the conduction and valence band deformation potentials.

\begin{figure}[t!]
	\centering
	\includegraphics*[width=\linewidth,height=0.9\linewidth]{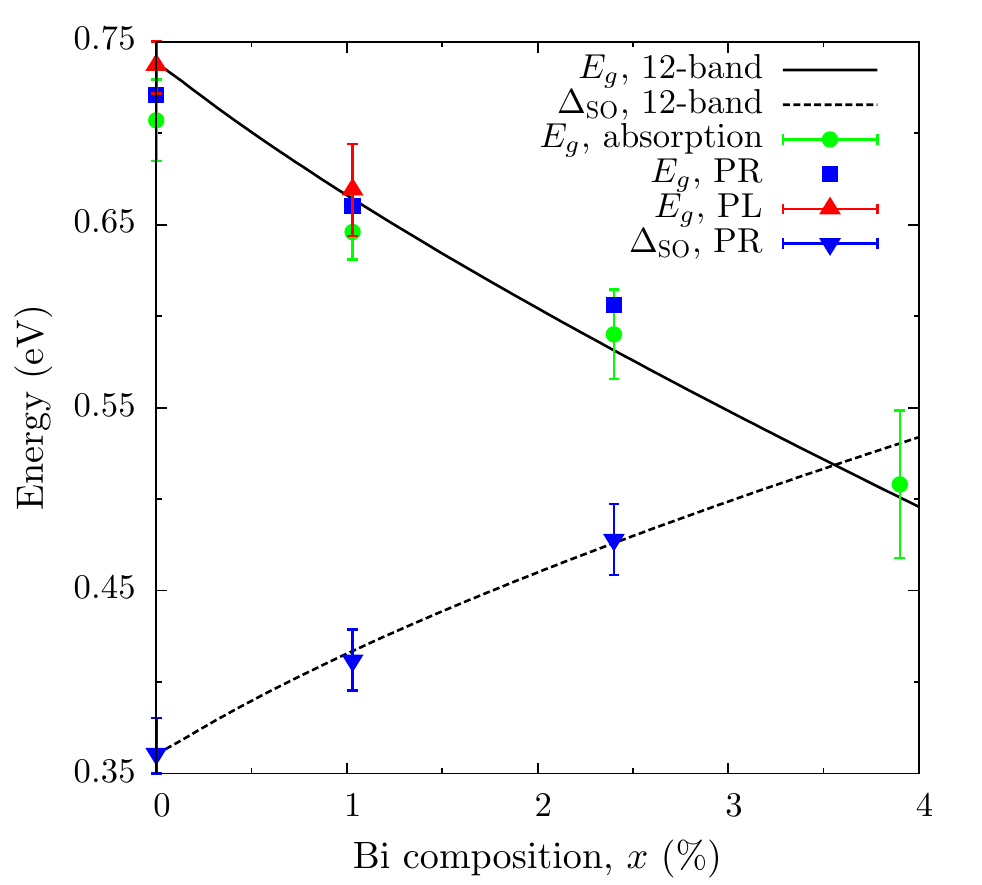}
	\caption{Measured and calculated variation of the band gap ($E_{g}$) and spin-orbit-splitting ($\Delta_{\protect\scalebox{0.7}{\textrm{SO}}}$) as a function of Bi composition ($x$) for In$_{0.53}$Ga$_{0.47}$Bi$_{x}$As$_{1-x}$ grown pseudomorphically on InP. Solid and dashed lines: Calculation of $E_{g}$ and $\Delta_{\protect\scalebox{0.7}{\textrm{SO}}}$ using the 12-band $\textbf{k}\cdot\textbf{p}$ model. Various symbols: Values of $E_{g}$ and $\Delta_{\protect\scalebox{0.7}{\textrm{SO}}}$ obtained from optical absorption, photo-modulated reflectance (PR) and photo-luminescence (PL) measurements on a series of MBE-grown In$_{0.53}$Ga$_{0.47}$Bi$_{x}$As$_{1-x}$/InP samples~\cite{Petropoulos_APL_2011,Marko_APL_2012,Broderick_SST_2012}.}
	\label{fig:InGaBiAs_energy_gaps}
\end{figure}

Figure~\ref{fig:InGaBiAs_energy_gaps} shows the calculated variation of $E_{g}$ and $\Delta_{\scalebox{0.7}{\textrm{SO}}}$ as a function of $x$ for In$_{0.53}$Ga$_{0.47}$Bi$_{x}$As$_{1-x}$ grown pseudomorphically on InP using the 12-band $\textbf{k}\cdot\textbf{p}$ model, compared to optical absorption, photo-modulated reflectance and photo-luminescence measurements. Details of the sample growth can be found in Ref.~\onlinecite{Petropoulos_APL_2011} and details of the measurements can be found in Ref.~\onlinecite{Marko_APL_2012}. Overall we note good agreement between the calculated and measured values throughout the investigated composition range. Based on the trends observed in the experimental and theoretical data, we conclude that the 12-band \textbf{k}$\cdot$\textbf{p} model provides a good description of the large band gap bowing observed in these samples.


We note also the excellent agreement between the calculated and measuredvalues of $\Delta_{\scalebox{0.7}{\textrm{SO}}}$ for the different compositions, confirming that the 12-band $\textbf{k}\cdot\textbf{p}$ Hamiltonian provides an accurate description of the band structure of dilute bismide alloys.


\section{Conclusions}

We have presented a 12-band $\textbf{k}\cdot\textbf{p}$ model for (In)GaBi$_{x}$As$_{1-x}$ alloys based on an $sp^{3}s^{*}$ tight-binding model of In$_{1-y}$Ga$_{y}$Bi$_{x}$As$_{1-x}$. The excellent agreement between the models confirms that the inclusion of band-anticrossing interactions in the valence band accurately describe the valence band structure of these dilute bismide alloys. The strong dependence of the calculated Bi-related impurity state energy ($E_{\scalebox{0.7}{\textrm{Bi}}}$) on supercell size using the $sp^{3}s^{*}$ tight binding method highlights the need to use large supercell calculations to analyse the band structure of (In)GaBi$_{x}$As$_{1-x}$. Our calculations of the band gap and spin-orbit-splitting energies are in good agreement with spectroscopic measurements on In$_{0.53}$Ga$_{0.47}$Bi$_{x}$As$_{1-x}$/InP samples, thus validating our models for further investigations of the electronic and optical properties of this interesting new class of semiconductor alloy.


\section*{Acknowledgements}

We would like to thank Prof. S. J. Sweeney of the University of Surrey for making available the results of experimental measurements prior to publication (in Ref.~\onlinecite{Marko_APL_2012}). C. A. Broderick acknowledges financial support from the Irish Research Council under the Embark Initiative (RS/2010/2766). E. P. O'Reilly and M. Usman acknowledge financial support from Science Foundation Ireland (10/IN.1/I299) and from the European Union Seventh Framework Programme under the project BIANCHO (FP7-257974).



\end{document}